\documentclass[11pt]{article}

\usepackage{amsmath,amssymb,graphicx}
\makeatletter

 \@addtoreset{equation}{section}
\makeatother

\newcommand{\Td}[2]{\frac{d{#1}}{d{#2}}}
\newcommand{\Pd}[2]{\frac{\partial{#1}}{\partial{#2}}}

\newcommand{\ltd}[1]{\frac{d}{d{#1}}}

\newcommand{\del}{\partial}

\newcommand{\half}{\frac{1}{2}}

\newcommand{\LS}{\ \ \ \ \ \ \ \ \ \ }

\newcommand{\ol}{\overline}

\newcommand{\kahler}{K\"{a}hler }

\newcommand{\bsubeq}{\begin{subequations}}
\newcommand{\esubeq}{\end{subequations}}
\newcommand{\vs}[1]{\vspace{#1 mm}}
\newcommand{\hs}[1]{\hspace{#1 mm}}

\newcommand {\beq}{\begin{eqnarray}}
\newcommand {\eeq}{\end{eqnarray}}
\newcommand {\non}{\nonumber\\}

\newcommand {\1}[1]{\frac{1}{#1}}

\newcommand {\theb}{\ol{\theta}}

\newcommand {\ph}{\varphi}

\newcommand {\sig}{\sigma}

\newcommand {\Ph}{\Phi}

\newcommand {\dagg}{^{\dagger}}

\newcommand {\lam}{\lambda}

\newcommand {\tr}{{\rm tr}\,}

\newcommand {\Lam}{\Lambda}

\begin{document}

\allowdisplaybreaks{

%%%%%%%%%%%%%%%%%%%%%%%%%%%%%%%%%%%%%%%%%%%%%%%%%%%%%%%
\setcounter{page}{0}

%%%%%%%%% title %%%%%%%%%%%%%
\begin{titlepage}

{\normalsize
\begin{flushright}
OU-HET 393\\
{\tt hep-th/0108084}\\
August 2001
\end{flushright}
}
\bigskip

\begin{center}
{\LARGE\bf Ricci-flat \kahler Manifolds\\

~from Supersymmetric Gauge Theories}

\vs{10}

\bigskip
{\renewcommand{\thefootnote}{\fnsymbol{footnote}}
{\Large\bf Kiyoshi Higashijima\footnote{
     E-mail: {\tt higashij@phys.sci.osaka-u.ac.jp}},
 Tetsuji Kimura\footnote{
     E-mail: {\tt t-kimura@het.phys.sci.osaka-u.ac.jp}} {\large and}
 Muneto Nitta\footnote{
     E-mail: {\tt nitta@het.phys.sci.osaka-u.ac.jp}}\footnote{
Address after September 1: Department of Physics, Purdue University,
West Lafayette, IN 47907-1396, USA.}
}}

\setcounter{footnote}{0}
\bigskip

{\large\sl
Department of Physics,
Graduate School of Science, Osaka University, \\
Toyonaka, Osaka 560-0043, Japan \\
}
\end{center}
\bigskip

%%%%%%%%% abstract %%%%%%%%

\begin{abstract}
Using techniques of supersymmetric gauge theories,
we present the Ricci-flat metrics on
non-compact \kahler manifolds whose
conical singularity is repaired
by the Hermitian symmetric space.
These manifolds can be identified as the complex line bundles
over the Hermitian symmetric spaces.
Each of the metrics contains a resolution parameter
which controls the size of these base manifolds,
and the conical singularity appears
when the parameter vanishes.

\end{abstract}

\end{titlepage}

%%%%%%%%%%%%%%%%%%%%%%%%%%%%%%%%%%%

\section{Introduction}
${\cal N}=2$ supersymmetric nonlinear sigma models
in two dimensions~\cite{Zu}
on Ricci-flat \kahler manifolds can be considered
as the model of the superstring theory
on curved space~\cite{finite,GVZ,NS}.
Ricci-flat \kahler manifolds are also
important ingredient for D-branes in curved space.
In the previous letter~\cite{HKN},
we presented the simple derivation of
the $O(N)$ symmetric Ricci-flat metric,
which actually coincides with the Stenzel metric on
the cotangent bundle over $S^{N-1}$~\cite{St}.
The conical singularity is resolved by $S^{N-1}$
with a radius being the deformation parameter.
It reduces to the Eguchi-Hanson gravitational instanton~\cite{EH}
and the six-dimensional deformed conifold~\cite{conifold}
in the cases of $N=3$ and $N=4$, respectively.
In \cite{PT}, a new metric for
the six-dimensional conifold, in which the conical singularity is
repaired by $S^2 \times S^2$, was found.
It was generalized in our previous letter \cite{HKN2}
to the higher dimensional conifold, in which
the singularity is resolved by the complex quadric surface
$Q^{N-2} = SO(N)/[SO(N-2)\times U(1)]$~\cite{HKN2}.
The new manifold can be regarded as
the complex line bundle over $Q^{N-2}$,
which is a Hermitian symmetric space.

In this paper, we present the new Ricci-flat metrics replacing
the base manifold by other Hermitian symmetric spaces,
the Grassmann manifold $G_{N,M}=SU(N)/[SU(N-M)\times U(M)]$,
$SO(2N)/U(N)$ and $Sp(N)/U(N)$.
To do this, we apply the technique of
the gauge theory formulation of
supersymmetric nonlinear sigma models
on the Hermitian symmetric spaces~\cite{HN1,HN2,HN3},
which was used for the study of non-perturbative effects~\cite{HKNT}.
We note that our manifolds are natural generalizations
of the Calabi metric on the complex line bundle
over ${\bf C}P^{N-1}$~\cite{Ca}.

This paper is organized as follows.
In Sec.~2, we recapitulate the construction
of compact \kahler manifolds
$G_{N,M}$, $SO(2N)/U(N)$ and $Sp(N)/U(N)$ by
supersymmetric gauge theories,
and extend this to non-compact \kahler manifolds.
In Sec.~3, we impose the Ricci-flat condition on
these non-compact manifolds.
Symmetry plays a crucial role to reduce partial differential
equations to ordinary differential equations of one variable.
In Sec.~4, we present explicit expressions of
\kahler metrics and their \kahler potentials.
It is found that these manifolds contain resolution parameter
$b$ as an integration constant,
and the conical singularity is resolved
by $G_{N,M}$, $SO(2N)/U(N)$ or $Sp(N)/U(N)$
of a radius expressed in terms of $b$.
These manifolds are complex line bundles over
$G_{N,M}$, $SO(2N)/U(N)$ and $Sp(N)/U(N)$.
Sec.~5 is devoted to Conclusion and Discussions.
In Appendix,
we summarize the isomorphisms between the lower dimensional
base manifolds and the duality between the Grassmann manifolds,
and show that they hold for total spaces.

\vs{20}

%%%%%%%%%%%%%%%%%%%%%%%%%%%%%%%%%%%
\section{Construction by Supersymmetric Gauge Theories}

%%%%%%%%%%%%%%%%%%%%%%%%%%%%%%%%%%%
\subsection{Compact \kahler Manifolds from Gauge Theories}
In this section we recapitulate
the construction of $G_{N,M}$, $SO(2N)/U(N)$ and $Sp(N)/U(N)$,
using supersymmetric gauge theories~\cite{HN1}.
Such a method was first found for the projective space
${\bf C}P^{N-1}$~\cite{DDL} and
the Grassmann manifold $G_{N,M}$~\cite{Ao},
and then recognized as the symplectic or
the \kahler quotient \cite{HKLR}.

%%%%%%%%%%%%%%%%
{\sl Construction of $G_{N,M}$}~\cite{Ao,HKLR}.
Let $\Phi(x,\theta,\theb)$ be an $N \times M$ matrix-valued
chiral superfield. The group $SU(N) \times U(M)$ can act on
it as
\begin{align}
 \Phi \to \Phi'
\ &= \ g_{\rm L} \Phi {g_{\rm R}}^{-1} \; , \ \ \
 (g_{\rm L},g_{\rm R}) \ \in \ (SU(N),U(M)) \; .
\end{align}
We promote the right action of $U(M)$ to a gauge symmetry
by introducing a vector superfield $V(x,\theta,\theb)$,
taking a value in the Lie algebra of $U(M)$.
The gauge transformation is given by
\begin{align}
 \Phi \to \Phi'
\ &= \ \Phi e^{-i\Lam} \; , \ \ \
 e^V \to e^{V'} \ = \ e^{i\Lam}e^V e^{-i\Lam^{\dagger}} \; ,
\end{align}
where $\Lam(x,\theta,\theb)$ is a parameter
chiral superfield, taking a value in the Lie algebra of $U(M)$.
Note that the local invariance group is enlarged to
the complexification of the gauge group,
$U(M)^{\bf C}=GL(N,{\bf C})$,
since the scalar component of
$\Lam(x,\theta,\theb)$ is complex.
The Lagrangian invariant under the global $SU(N)$ and
the local $U(M)$ symmetries is given by
\begin{align}
  {\cal L}
\ &= \ \int \! d^4 \theta \, {\cal K} (\Phi,\Phi^{\dagger},V)
\ = \ \int \! d^4 \theta
    \left[ \tr(\Phi^{\dagger}\Phi e^V) - c\, \tr V \right] \; ,
     \label{Gras.}
\end{align}
where ${\cal K}$ is the \kahler potential.
Here $c$ is a real positive constant, called the
Fayet-Iliopoulous (FI) parameter,
and $c\, \tr V$ is called the FI D-term.

Since $V$ is an auxiliary field, it can be eliminated by
its equation of motion~\footnote{
We regard $e^{-V}\delta e^V$ as an infinitesimal
parameter:
$\delta {\cal L} =
\tr [\Ph\dagg\Ph e^V (e^{-V}\delta e^V)]
- c \, \tr (\delta \log e^V )
= \tr [(\Ph\dagg\Ph e^V - c {\bf 1}_M) X^{-1}\delta X]$,
where $X = e^V$.
}
\begin{align}
 \delta {\cal L} / \delta V \ &= \ \Ph\dagg\Ph e^V - c {\bf 1}_M \ = \
 0 \; ,
\end{align}
where ${\bf 1}_M$ is an $M \times M$ unit matrix.
Substituting the solution,
$V(\Ph,\Ph\dagg) = - \log \left({\Ph\dagg\Ph/ c}\right)$,
back into the Lagrangian~(\ref{Gras.}),
we obtain
\begin{align}
 {\cal K} (\Ph,\Ph\dagg,V(\Ph,\Ph\dagg))
\ &= \ c\,\tr \log (\Ph\dagg\Ph)
\ =  \ c \log \det (\Ph\dagg\Ph) \non
\ &= \ c \log \det ({\bf 1}_M + \ph^{\dagger} \ph ) \;
. \label{Gra-Kahler}
\end{align}
Since the gauge group is complexified,
we have chosen the gauge fixing as
\begin{align}
 \Phi \ &= \ \begin{pmatrix}
        {\bf 1}_M \cr
                  \ph
        \end{pmatrix} \; ,  \label{fix}
\end{align}
where $\ph(x,\theta,\theb)$ is an $(N-M)\times M$
matrix-valued chiral superfield.
The constant terms in (\ref{Gra-Kahler}) have been omitted,
since they disappear under the superspace integral $\int d^4\theta$.
(\ref{Gra-Kahler}) is the \kahler potential of
$G_{N,M}=SU(N)/[SU(N-M)\times U(M)]$,
whose complex dimension is $M(N-M)$.
It becomes one of ${\bf C}P^{N-1}=SU(N)/[SU(N-1)\times U(1)]$
if we set $M=1$,
in which case the gauge group is $U(1)$.

%%%%%%%%%%
{\sl Construction of $SO(2N)/U(N)$ and $Sp(N)/U(N)$}~\cite{HN1}.
Let us replace the size of the matrix $\Phi$,
$(N,M)$, by $(2N,N)$, corresponding to $G_{2N,N}$,
and introduce the invariant tensor of $SO(2N)$ or $Sp(N)$:
\begin{align}
 J \ &= \ \begin{pmatrix}
              {\bf 0} & {\bf 1}_N \cr
   \epsilon{\bf 1}_N  & {\bf 0}
 \end{pmatrix} \; , \label{J}
\end{align}
where $\epsilon = 1$ for $SO(2N)$, and $\epsilon = -1$ for $Sp(N)$.
The invariant Lagrangian is given by
\begin{align}
{\cal L}
\ &= \ \int \! d^4 \theta
  \left[ \tr(\Phi^{\dagger}\Phi e^V) - c\, \tr V \right]
 + \left[ \int d^2 \theta \, \tr (\Phi_0 \Phi^T J \Phi)
 + {\rm c.c} \right] \; , \label{So-Sp}
\end{align}
where $\Phi_0(x,\theta,\theb)$ is
an auxiliary chiral superfield of an $N \times N$ matrix,
belonging to (anti-)symmetric tensor representation of
the gauge group $U(N)$ for $SO(2N)/U(N)$ [$Sp(N)/U(N)$]
with the suitable $U(1)$ charge.

By the integration over $V$, we obtain (\ref{Gra-Kahler})
with the same gauge fixing as (\ref{fix}).
The integration over $\Phi_0$ gives the constraint
\begin{align}
 \Phi^T J \Phi \ &= \ \ph + \epsilon \ph^T  \ = \ 0 \; ,
 \label{const-SO-Sp}
\end{align}
which implies that the $N \times N$ matrix-valued chiral superfield
$\ph$ is anti-symmetric or symmetric
for $SO(2N)/U(N)$ or $Sp(N)/U(N)$, respectively.
The \kahler potential (\ref{Gra-Kahler}) with
the constraints (\ref{const-SO-Sp}) is
one of $SO(2N)/U(N)$ or $Sp(N)/U(N)$,
whose complex dimension is $\1{2}N(N-1)$ or
$\1{2}N(N+1)$, respectively.

Instead of the \kahler potential of
the Lagrangian (\ref{Gras.}) and (\ref{So-Sp}),
we can start from
\begin{align}
  {\cal K}(\Phi,\Phi^{\dagger},V) \ &= \
    f(\tr(\Phi^{\dagger}\Phi e^V)) - c\, \tr V  \; ,
     \label{arbitrary-k}
\end{align}
where $f$ is an {\it arbitrary} function.
We can show that we obtain the same results
even if we start from (\ref{arbitrary-k})~\cite{HN2}.
Let us make some comments.
We have used the classical equation of motion of $V$
to eliminate it.
We can promote this to the quantum level in
the path integral formalism~\cite{HN2}.
If we add the kinetic term for $V$ rather than regarding $V$
as auxiliary, our manifolds are obtained as
the classical moduli space of the gauge theories~\cite{LT}.

%%%%%%%%%%%%%%%%%%%%%%%%%%%%%%%%%%%
\subsection{Non-compact \kahler Manifolds from Gauge Theories}

Let us construct the non-compact \kahler manifolds,
by restricting the gauge degrees of
freedom from $U(M)$ to $SU(M)$.
To do this, we promote the FI-parameter $c$
in (\ref{arbitrary-k})
to an auxiliary vector superfield $C(x,\theta,\theb)$:
\begin{align}
{\cal K}_0 (\Phi, \Phi^{\dagger}, V, C)
 \ = \  f ({\rm tr}(\Phi^{\dagger} \Phi e^V))
       - C \, {\rm tr} V \; , \label{potential}
\end{align}
where $f$ is an arbitrary function.\footnote{
There exist independent invariants
${\rm tr}[(\Phi^{\dagger} \Phi e^V)^2]$, $\cdots$,
${\rm tr}[(\Phi^{\dagger} \Phi e^V)^M]$,
besides ${\rm tr}(\Phi^{\dagger} \Phi e^V)$.
We can show that, even if these are included as the arguments of
the arbitrary function of (\ref{potential}),
we obtain the same result (\ref{kahler-potential}).
The situation is the same for
the cases of the $U(M)$ gauge field, (\ref{arbitrary-k}),
for compact manifolds.
}
Note that $V(x,\theta,\theb)$ in this Lagrangian
is still taking a value in the Lie algebra of $U(M)$.
The equations of motion of $V$ and $C$ read
\bsubeq
\begin{align}
 & \delta {\cal L}/\delta V
   \ = \ f'({\rm tr}(\Phi^{\dagger} \Phi e^V))\,
   \Phi^{\dagger} \Phi e^V
   - C {\bf 1}_M \ = \ 0 \; , \label{EOM-V} \\
 & \delta {\cal L}/\delta C \ = \ \tr V \ = \ 0 \; ,\label{EOM-C}
\end{align}
\esubeq
respectively, where the prime denotes the differentiation with respect
to the argument of $f$.
The gauge group is restricted to $SU(M)$ by (\ref{EOM-C}).
The trace and the determinant of (\ref{EOM-V}) are
\bsubeq
\begin{align}
 f'({\rm tr}(\Phi^{\dagger} \Phi e^V))\,
   \tr(\Phi^{\dagger} \Phi e^V) \ &= \ M C \;, \\
 \left[f'({\rm tr}(\Phi^{\dagger} \Phi e^V)) \right]^M
 \det (\Phi\dagg\Phi) \ &= \ C^M \;,
\end{align}
\esubeq
since $\det e^V=1$ for the $SU(M)$ gauge field $V$.
Eliminating $C$ from these equations,
the solution of $V$ reads
\beq
 \tr (\Phi\dagg\Phi e^V) \ = \
 M \left[\det (\Phi\dagg\Phi) \right]^{\1{M}} \; .
\eeq
Substituting this back into (\ref{potential})
and taking account of (\ref{EOM-C}),
we obtain the nonlinear \kahler potential
\begin{align}
 {\cal K}_0 (\Phi, \Phi^{\dagger},V(\Phi,\Phi\dagg))
 \ = \ f \left(M \left[\det (\Phi\dagg\Phi) \right]^{\1{M}}\right)
 \ \equiv \ {\cal K} (X(\Phi, \Phi^{\dagger})) \; ,
  \label{kahler-potential}
\end{align}
where $X(\Phi, \Phi^{\dagger})$ is a vector superfield,
invariant under the global $U(N)$ [$SO(2N)$ or $Sp(N)$]
and the local $SU(M)$ [$SU(N)$] symmetries,
defined by
\begin{align}
 X(\Phi, \Phi^{\dagger}) \ &= \ \log \det \Phi^{\dagger} \Phi \; ,
\end{align}
and ${\cal K}(X)$ is a real function of $X$
related to $f$.
Here the logarithm in the definition of $X$
is just a convention. (Note that this definition of
the invariant is different from the one in \cite{HKN,HKN2}.) From
the view point of the algebraic variety,
$X$ is the gauge invariant parameterizing the moduli space
of supersymmetric gauge theories~\cite{LT}.

Since the gauge group is complexified to
$SU(M)^{\bf C}=SL(M,{\bf C})$,
we can choose a gauge fixing as
\begin{align}
\Phi \ &= \ \sig \begin{pmatrix}
  {\bf 1}_M \cr
  \varphi
 \end{pmatrix} \; , \label{scalar}
\end{align}
where $\varphi$ is an $(N-M) \times M$ matrix-valued chiral superfield,
and $\sig(x,\theta,\theb)$ is a chiral superfield.
Comparing (\ref{scalar}) with (\ref{fix}),
we find that the superfield $\sig$ is parameterizing a fiber,
while $\ph$ is parameterizing a base manifold,
with the total space being a complex line bundle.
Under this gauge fixing, the invariant superfield
$X$ is decomposed as
\begin{align}
X \ &= \ M \log |\sig|^2
  + \log \det ({\bf 1}_M + \varphi^{\dagger}\varphi)
\ = \  M \log |\sig|^2 + \Psi \; ,  \label{parameter}
\end{align}
where we have defined
\begin{align}
 \Psi \ &\equiv \
 \log \det ({\bf 1}_M + \varphi^{\dagger}\varphi)
 \; . \label{Psi}
\end{align}
Note that $\Psi$ is a \kahler potential of
$G_{N,M}$ [$SO(2N)/U(N)$ or $Sp(N)/U(N)$] obtained
in (\ref{Gra-Kahler})
[with the constraint (\ref{const-SO-Sp})].

Let us introduce some notations.
We denote the elements of the matrix-valued chiral superfield
$\ph$ by $\varphi_{A a}$,
where the upper case and the lower case indices,
$A$ and $a$, run from $1$ to $N-M$ and from $1$ to $M$, respectively.
Since the size of the matrix $\varphi$
is $N \times N$ in the cases of $Sp(N)/U(N)$ and $SO(2N)/U(N)$,
we denote its elements by $\ph_{ab}$.
In this case, only the components $\ph_{ab}$ with
$b \geq a$ ($b > a$) are considered as independent.
When we discuss the total space,
we use the coordinates $z^{\mu} \equiv (\sig, \varphi_{A a})$.
It should be noted that from now on
we use the {\it same} letters for chiral superfields {\it and}
their complex scalar components.

We make a comment on the symmetry breaking.
These non-compact manifolds can be regarded as
\beq
 {\bf R}\times {G\over H} \ = \
 {\bf R}\times {SU(N) \over SU(N-M) \times SU(M)}, \ \ \
 {\bf R}\times {SO(2N) \over SU(N)}, \ \ \
 {\bf R}\times {Sp(N) \over SU(N)} \;,
\eeq
at least locally.
The part of $G/H$ is parametrized by
the Nambu-Goldstone bosons arising from
the spontaneous breaking of the global symmetry $G$ down to $H$,
whereas the factor of ${\bf R}$ is parametrized by
the so-called quasi-Nambu-Goldstone boson
(see e.g. \cite{NLR,Ni1,HN1}).

%%%%%%%%%%%%%%%%%%%%%%%%%%%%%%
\section{Ricci-flat Conditions}

We would like to determine the function ${\cal K}(X)$
in (\ref{kahler-potential}),
by imposing the Ricci-flat condition on the manifold.
The metric of the \kahler manifold is given by
$g_{\mu \nu^*} = \del_{\mu} \del_{\nu^*} {\cal K}$,
where $\del_{\mu} = \del/ \del z^{\mu}$ and
$\del_{\nu^*} = \del / \del z^{* \nu}$.
The explicit expression of the \kahler metric can be
calculated as
\bsubeq
\label{metric}
\begin{align}
 g_{\mu \nu^*} \ &= \ \left(
  \begin{array}{cc}
        g_{\sig \sig^*}  & g_{\sig (B b)^*} \\
        g_{(A a) \sig^*} & g_{(A a) (B b)^*}
\end{array} \right) \; ,
\end{align}
with each block being
\begin{align}
 & g_{\sig \sig^*} \ = \ {\cal K}''
  \Pd{X}{\sig} \Pd{X}{\sig^*} \; , \hspace{5mm}
 g_{(A a) (B b)^*} \ = \
  {\cal K}'' \Pd{X}{\varphi_{A a}} \Pd{X}{\varphi^*_{B b}}
  + {\cal K}' {\partial^2 X \over
       \partial \varphi_{A a} \partial \varphi^*_{B b}} \; , \\
 & g_{\sig (B b)^*} \ = \
  {\cal K}''  \Pd{X}{\sig} \Pd{X}{\varphi^*_{B b}} \;,
 \hspace{5mm}
 g_{(A a) \sig^*} \ = \
  {\cal K}'' \Pd{X}{\varphi_{A a}} \Pd{X}{\sig^*} \; ,
\end{align}
\esubeq
where the prime denotes the differentiation with respect
to the argument $X$ of the function ${\cal K}(X)$.
Here, we have used equations,
${\partial^2 X \over \partial \sig \partial \sig^*}
= {\partial^2 X \over \partial \sig \partial \varphi^*_{B b}}
= {\partial^2 X \over \partial \varphi_{A a} \partial \sig^*}= 0$
($\sig \neq 0$), which follow from (\ref{parameter}).

The determinant of the metric is calculated, to yield
\begin{align}
\det g_{\mu \nu^*}
\ &= \ \frac{M^2}{|\sig|^2} {\cal K}'' \cdot
 \det_{(A a) (B b)^*} \left( {\cal K}'
  {\partial^2 X \over
   \partial \varphi_{A a} \partial \varphi^*_{B b} } \right)
\; , \label{determinant}
\end{align}
where $\det_{(A a) (B b)^*}$ denotes the determinant
of the matrix of the tensor product,
spanned by $(A a)$ and $(B b)^*$.
Since the Ricci-form is given by $(Ric)_{\mu \nu^*}
= - \del_{\mu} \del_{\nu^*} \log \det g_{\kappa\lam^*}$,
the Ricci-flat condition $(Ric)_{\mu \nu^*} = 0$ implies
\beq
 \det g_{\mu \nu^*}  =  \mbox{(constant)} \times |F|^2 \;,
 \label{ricci-flat}
\eeq
with $F$ being a holomorphic function.

%%%%%%%%%%%%%%%%%%%%%
\subsection{Line Bundle over $G_{N,M}$}

In this section let us obtain the explicit solution of
the Ricci-flat metric on the line bundle over
the Grassmann manifold.
Let us calculate the $X$ differentiated by
matrix fields $\ph_{A a}$ once and twice,
needed for the calculation of the \kahler metric (\ref{metric}).
By noting
\begin{align}
 \frac{\del \varphi_{B b}}{\del \varphi_{A a}} \
   &= \ \delta_{A B} \delta_{a b} \; , \label{deriv}
\end{align}
we obtain
\bsubeq
\label{deriv-G}
\begin{align}
 {\partial X \over \partial \varphi_{A a}}
 \ &= \ \del_{(Aa)}\Psi
 \ = \ \Big[({\bf 1}_M + \varphi^{\dagger} \varphi)^{-1}
             \ph^{\dagger}\Big]_{aA} \; , \\
 {\partial X \over \partial \varphi^*_{A a} }
 \ &= \ \del_{(Aa)^*}\Psi
 \ = \ \Big[\ph ({\bf 1}_M + \varphi^{\dagger} \varphi)^{-1}
            \Big]_{Aa} \; ,\\
 {\partial^2 X \over \partial \varphi_{A a} \partial \varphi^*_{B b}}
 \ &= \ \del_{(Aa)}\del_{(Bb)^*}\Psi
 \ = \ ({\bf 1}_M + \varphi^{\dagger} \varphi)^{-1}_{a b}
 \Big[ {\bf 1}_{(N-M)} -
   \varphi ({\bf 1}_M + \varphi^{\dagger} \varphi)^{-1}
   \varphi^{\dagger} \Big]_{B A} \;. \label{2-deriv}
\end{align}
\esubeq
Here we have used the definition of $\Psi$ in (\ref{Psi}),
and $\del_{(Aa)}$ and $\del_{(Aa)^*}$ represent
differentiations with respect to $\ph_{Aa}$ and $\ph_{Aa}^*$,
respectively.
Note that (\ref{2-deriv}) is just the \kahler metric of
the Grassmann manifold $G_{N,M}$.
The determinant of $g_{\mu \nu^*}$ can be calculated as
\begin{align}
 \det g_{\mu \nu^*} \ &= \ \frac{M^2}{|\sig|^2}
     {\cal K}'' ({\cal K}')^{M(N-M)} \cdot
 \det_{(A a) (B b)^*}
   \left[  \del_{(Aa)}\del_{(Bb)^*}\Psi \right] \; .
 \label{det-g}
\end{align}
To obtain the concrete expression of this determinant,
we use a symmetry transformation preserving
the value of the determinant.
Under the transformation of the complex isotropy
$[SU(N-M)\times SU(M)]^{\bf C}
= SL(N-M,{\bf C})\times SL(M,{\bf C})$,
the coordinates transform linearly as
$z^{\mu} \to z^{\prime \mu} =
{V^{\mu}}_{\nu} z^{\nu}$.
Since the transformation matrix $V$ belongs to
a subgroup of $SL(M(N-M),{\bf C})$,
the equation $\det V=1$ holds and
the $\det g_{\mu \nu^*}$ is invariant:
$\det g_{\mu \nu^*} \to \det g_{\mu \nu^*}' = \det g_{\mu
\nu^*} |\det V|^2 = \det g_{\mu \nu^*}$.
For an arbitrary matrix-valued chiral superfield $\ph$,
there exists a complex isotropy which
permits the transformation of $\ph$ to the form of
\begin{align}
\varphi \ &= \ \left(
\begin{array}{cc}
    \varphi_0 &  \dots \\
       \vdots & {\bf 0}
\end{array} \right) \; , \label{special}
\end{align}
where the dots denote zero elements,
and the only non-zero element is $\varphi_{11} \equiv \varphi_0$.
The matrix of the tensor product (\ref{2-deriv}) is diagonalized as
\begin{align}
 {\partial^2 X \over \partial \varphi_{A a} \partial \varphi^*_{B b}}
 \ &= \ {\rm diag.}\;
 \Big(\xi^2, \underbrace{\xi,\cdots,\xi}_{M-1}\,;\,
 \overbrace{
  \xi, \underbrace{1,\cdots,1}_{M-1} \,;\,  \cdots \,;\,
  \xi, \underbrace{1,\cdots,1}_{M-1}
 }^{\mbox{$(N-M-1)$-blocks}} \Big) \; ,
\end{align}
where $\xi \equiv (1 +|\ph_0|^2)^{-1}$.
Each block separated by the semicolons is labeled by
the indices $A=B$, which run from $1$ to $N-M$,
and in each block the indices $a=b$ run from $1$ to $M$.
With noting $\xi = (1 +|\ph_0|^2)^{-1}
= [\det({\bf 1}_N + \ph^{\dagger}\ph)]^{-1}
= |\sig|^{2M} e^{-X}$,
the determinant (\ref{det-g}) can be calculated as
\begin{align}
\det g_{\mu \nu^*} \ &= \ M^2 |\sig|^{2(MN-1)} e^{-N X} {\cal K}''
            ({\cal K}')^{M(N-M)} \; . \label{det-g-2}
\end{align}
The Ricci-flat condition (\ref{ricci-flat}) becomes
\begin{align}
 e^{-N X} \ltd{X} ({\cal K}')^{M(N-M)+1} \ &= \ a \; , \label{ODE-G}
\end{align}
where $a$ is a real constant.

%%%%%%%%%%%%%%%%%%%%%%%%%%%%%%%%%%%%%%%%%%
\subsection{Line Bundles over $SO(2N)/U(N)$ and $Sp(N)/U(N)$}

In this section, we construct the Ricci-flat metrics on
the line bundles over $SO(2N)/U(N)$ and $Sp(N)/U(N)$.
These cases are obtained by imposing the constraint
(\ref{const-SO-Sp}) on the Grassmann manifold $G_{2N, N}$.
Under the condition (\ref{const-SO-Sp}),
the differentiations with respect to
the matrix elements $\varphi_{ab}$,
corresponding to (\ref{deriv}) for the Grassmann case, become
\begin{align}
 {\partial\varphi_{c d} \over \partial\varphi_{a b}}
 \ &= \ \big( \delta_{c a} \delta_{d b}
   - \epsilon \; \delta_{c b} \delta_{d a} \big)
   \Big( 1 - \half \delta_{a b} \Big) \; ,
\end{align}
where we do not take a sum over the index $a$ or $b$.
Using this, the $X$ differentiated by one or two $\ph$'s
can be calculated, to yield
\bsubeq
 \label{deriv-SO-Sp}
\begin{align}
{\partial X \over \partial \varphi_{ab}}
 \ &= \ \del_{(ab)}\Psi
 \ = \ \sum_{c, d=1}^N
   \Big[({\bf 1}_N + \varphi^{\dagger} \varphi)^{-1}
             \ph^{\dagger}\Big]_{dc}
   \big( \delta_{c a} \delta_{d b}
   - \epsilon \; \delta_{c b} \delta_{d a} \big)
   \Big( 1 - \half \delta_{a b} \Big) \; , \\
 {\partial X \over \partial \varphi^*_{ab} }
 \ &= \ \del_{(ab)^*}\Psi
 \ = \ \sum_{c, d=1}^N
  \Big[\ph ({\bf 1}_N + \varphi^{\dagger}
                  \varphi)^{-1} \Big]_{cd}
     \big( \delta_{c a} \delta_{d b}
   - \epsilon \; \delta_{c b} \delta_{d a} \big)
   \Big( 1 - \half \delta_{a b} \Big)  \; ,\\
 {\partial^2 X \over \partial\varphi_{a b}
   \partial\varphi^*_{c d}}
 \ &= \ \del_{(ab)} \del_{(cd)^*} \Psi \non
 \ &= \ \Big( 1 - \half \delta_{a b} \Big)
  \Big( 1 - \half \delta_{c d} \Big)
  \Big\{ ({\bf 1}_N + \varphi^{\dagger} \varphi)^{-1}_{b d}
   \big[ {\bf 1}_N
     - \varphi ({\bf 1}_N + \varphi^{\dagger}\varphi)^{-1}
       \varphi^{\dagger} \big]_{c a} \nonumber \\
 \ & \LS \ \ \ \
 - \epsilon \; ({\bf 1}_N + \varphi^{\dagger}\varphi)^{-1}_{b c}
   \big[ {\bf 1}_N
      - \varphi ({\bf 1}_N + \varphi^{\dagger} \varphi)^{-1}
        \varphi^{\dagger} \big]_{d a}
 + (a \leftrightarrow b, c \leftrightarrow d)
 \Big\} \;.
  \label{2-deriv-so-sp}
\end{align}
\esubeq
Here the last term in the last line
implies adding the preceding two terms
with the exchange of the indices.
Note again that (\ref{2-deriv-so-sp}) is just
the \kahler metric of $SO(2N)/U(N)$ or $Sp(N)/U(N)$.
The determinant (\ref{determinant}) can be calculated as
\begin{align}
 \det g_{\mu \nu^*} \ &= \ \frac{N^2}{|\sig|^2} {\cal K}'' \,
  ({\cal K}')^{\1{2}N(N - \epsilon)} \cdot
  \det_{(a b) (c d)^*}
  \left[\del_{(ab)} \del_{(cd)^*} \Psi \right]\;.
 \label{det-so-sp}
\end{align}
We again use the complex isotropy transformation of
$SU(N)^{\bf C} = SL(N,{\bf C})$, preserving the determinant.
We first discuss $SO(2N)/U(N)$ followed by $Sp(N)/U(N)$.

%%%%%%%%%%%
{\sl The line bundle over $SO(2N)/U(N)$.}
Using the complex isotropy transformation of $SL(N,{\bf C})$,
the arbitrary $\ph$ can be put
\begin{align}
\varphi \ &= \ \left(
\begin{array}{ccc}
          0 & \varphi_0 & 0 \\
 -\varphi_0 & 0         & 0 \\
          0 & 0         & {\bf 0}
\end{array} \right) \ ,
\end{align}
where non-zero elements are $\ph_{12} = -\ph_{21} \equiv \ph_0$.
The matrix of the tensor product (\ref{2-deriv-so-sp})
is diagonalized as
\begin{align}
 \left. \frac{\del^2 X}{\del \varphi_{a b}
    \del \varphi^*_{c d}}\right|_{b> a,\, d> c}
\ &= \ {\rm diag.}\,
 \Big(2 \xi^2, \underbrace{2\xi,\cdots,2\xi}_{N-2}\,;\,
 \underbrace{2\xi,\cdots,2\xi}_{N-2}\,;\,
 \underbrace{2,\cdots,2}_{N-3}\,;\,
 \underbrace{2,\cdots,2}_{N-4}\,;\,
 \cdots \,;\, \underbrace{2,2}_2 \,;\, 2 \Big) \; ,
\end{align}
where $\xi \equiv (1+|\ph_0|^2)^{-1}$.
Each block separated by the semicolons is labeled by
the indices $a=c$, which run from $1$ to $N$,
and the indices $b=d$ run from $a+1=c+1$ to $N$ in the $a$-th block,
by the conditions $b > a$ and $d > c$.
Noting $\xi = (1+|\ph_0|^2)^{-1}
= [\det({\bf 1}_N + \ph^{\dagger}\ph)]^{-1/2}
= |\sig|^N e^{-X/2}$,
we can calculate the determinant (\ref{det-so-sp}), given by
\begin{align}
\det g_{\mu \nu^*} \ &= \ N^2 2^{\half N(N-1)} |\sig|^{2N(N-1)-2}
   e^{-(N-1) X} \, {\cal K}'' ({\cal K}')^{\half N(N-1)} \; .
   \label{det-so}
\end{align}
The Ricci-flat condition (\ref{ricci-flat}) becomes
\begin{align}
  e^{-(N-1) X} \ltd{X}
  ({\cal K}')^{\half N(N-1) + 1} \ &= \ a \; . \label{ODE-SO}
\end{align}

%%%%%%%%%
{\sl The line bundle over $Sp(N)/U(N)$.}
There exists an isotropy transformation which
transforms an arbitrary matrix $\varphi$
to the form of (\ref{special}).
The matrix of the tensor product (\ref{2-deriv-so-sp})
is diagonalized as
\begin{align}
 \left. \frac{\del^2 X}{\del \varphi_{a b}
  \del \varphi^*_{c d}} \right|_{b\geq a,\, d \geq c}
  \ &= \ {\rm diag.}\;
 \Big(\xi^2, \underbrace{2\xi,\cdots,2\xi}_{N-1}\,;\,
 1,\underbrace{2,\cdots,2}_{N-2}\,;\,
 1,\underbrace{2,\cdots,2}_{N-3}\,;\,
 \cdots \,;\, 1, 2 \,;\, 1 \Big) \; ,
\end{align}
where $\xi \equiv (1 +|\ph_0|^2)^{-1}$.
Each block separated by the semicolons is labeled by
the indices $a=c$, which run from $1$ to $N$,
and the indices $b=d$ run from $a=c$ to $N$ in the $a$-th block
by the conditions $b \geq a$ and $d \geq c$.
Noting $\xi = (1 +|\ph_0|^2)^{-1}
= [\det({\bf 1}_N + \ph^{\dagger}\ph)]^{-1}
= |\sig|^{2N} e^{ -X}$,
we can calculate the determinant (\ref{det-so-sp}), given by
\begin{align}
 \det g_{\mu \nu^*}  \ &= \ N^2 2^{\half N(N-1)} |\sig|^{2N(N+1)-2}
    e^{-(N+1) X} \,{\cal K}'' ({\cal K}')^{\half N(N+1)}  \; .
\end{align}
The Ricci-flat condition (\ref{ricci-flat}) becomes
\begin{align}
 e^{-(N+1) X} \ltd{X}
 ({\cal K}')^{\half N(N+1) + 1} \ &= \ a \; . \label{ODE-Sp}
\end{align}

%%%%%%%%%%%%%%%%%%%%%%%%%%%%%%%%%%%%%%%%%%%%%%%
\section{Ricci-flat Metrics and \kahler Potentials}

\subsection{\kahler Potentials}

We can immediately solve (\ref{ODE-G}),
(\ref{ODE-SO}) and (\ref{ODE-Sp}):
\begin{align}\label{deriv-K}
\Td{{\cal K}}{X} \ = \
\left\{
 \begin{array}{l}
 \big( \lambda e^{N X} + b \big)^{\frac{1}{g}}
  \; , \ \ \ \ \ \ \ \ g \ \equiv \ M(N-M) +1 \;
   \ \ \mbox{ for $G_{N,M}$} \ ,\\
 \big( \lambda e^{(N-1) X} + b \big)^{\frac{1}{f}}
  \; , \ \ \ f \ \equiv \ \half N (N-1) + 1 \;
   \ \ \mbox{ for $SO(2N)/U(N)$} \ , \\
 \big( \lambda e^{(N+1) X} + b \big)^{\frac{1}{h}}
 \; , \ \ \ h \ \equiv \ \half N (N+1) + 1 \;
   \ \ \mbox{ for $Sp(N)/U(N)$} \ ,
\end{array}
\right.
\end{align}
where $\lam$ is a constant related to $a$, $N$ and $M$,
and $b$ is an integration constant
interpreted as a {\it resolution parameter} of
the conical singularity.
Although these are sufficient to obtain the \kahler metrics
using (\ref{metric}),
we can calculate \kahler potentials themselves:
\begin{align}\label{LG-kahler}
{\cal K}(X) \ = \
\left\{
 \begin{array}{l}
 {g \over N}
  \left[ (\lam e^{NX} + b)^{\1{g}} + b^{\1{g}}
  \cdot I \big( b^{-\1{g}} (\lam e^{NX} + b)^{\1{g}}
               ; g \big) \right] \;
  \ \ \hs{17} \mbox{ for $G_{N,M}$} \ ,\\
 \frac{f}{N-1}
  \left[ (\lam e^{(N-1)X} + b)^{\frac{1}{f}}
   + b^{\frac{1}{f}} \cdot
   I \big( b^{-\1{f}} (\lam e^{(N-1)X} + b)^{\frac{1}{f}}
        ;f\big) \right] \;
   \ \ \mbox{ for $SO(2N)/U(N)$} \ , \\
 \frac{h}{N+1}
  \left[ (\lam e^{(N+1)X} + b)^{\frac{1}{h}} + b^{\frac{1}{h}}
  \cdot I \big( b^{-\1{h}} (\lam e^{(N+1)X} + b)^{\frac{1}{h}}
  ; h \big) \right] \;
   \ \ \mbox{ for $Sp(N)/U(N)$} \ .
\end{array}
\right.
\end{align}
Here the function $I(y;n)$ is defined by
\begin{align}
I (y; n)
\ \equiv \
  \int^{y} \! \frac{dt}{t^n - 1}
\ &= \ \frac{1}{n} \Big[ \log \big( y - 1 \big)
    - \frac{1 + (-1)^n}{2}
    \log \big( y + 1 \big) \Big] \nonumber \\
& \ \ \ \
 + \frac{1}{n} \sum_{r=1}^{[\frac{n-1}{2}]} \cos \frac{2 r \pi}{n}
\cdot \log \Big( y{}^2 - 2 y \cos \frac{2 r \pi}{n} + 1 \Big)
\nonumber \\
\ & \ \ \ \ + \frac{2}{n} \sum_{r=1}^{[\frac{n-1}{2}]}
\sin \frac{2 r \pi}{n}
\cdot \arctan \Big[ \frac{\cos (2 r \pi / n) - y}
                    {\sin (2 r \pi /n) } \Big] \; .
\end{align}

In the limit of $b\to 0$, these manifolds become
(generalized) conifolds with their \kahler potentials,
\begin{align}
{\cal K} \ = \
\left\{ \begin{array}{l}
 {g \lam^{\1{g}} \over N} \ \big[|\sig|^{2M}
  \det( {\bf 1}_M + \ph^{\dagger} \ph) \big]^{N \over g}\;
  \ \ \hs{28.5} \mbox{ for $G_{N,M}$} \ ,\\
 {f \lam^{\1{f}} \over N-1} \ \big[|\sig|^{2N}
  \det ({\bf 1}_N + \ph^{\dagger} \ph) \big]^{N-1 \over f}
  \ , \hspace{0.5cm} \ph^T = - \ph \;,
  \ \ \mbox{ for $SO(2N)/U(N)$} \ , \\
 {h \lam^{\1{h}} \over N+1} \ \big[|\sig|^{2N}
  \det ({\bf 1}_N + \ph^{\dagger} \ph) \big]^{N+1 \over h}
   \ , \hs{5} \ph^T = \ph \; ,
   \ \ \ \ \, \mbox{ for $Sp(N)/U(N)$} \ .
\end{array}
\right.
\end{align}

%%%%%%%%%%%
\subsection{Ricci-flat \kahler Metrics}
We can calculate the Ricci-flat \kahler metrics
substituting the solutions  (\ref{deriv-K}) into (\ref{metric}).
The component $g_{\sig\sig^*}$ is
\begin{align}
g_{\sigma \sigma^*} \ = \
 \left\{
  \begin{array}{l}
 \lambda \frac{M^2 N}{g}
  \big( \lambda e^{NX} + b \big)^{\frac{1}{g} - 1}
  e^{N \Psi} |\sigma|^{2 MN - 2} \; ,
   \ \ \hs{23.9} \mbox{ for $G_{N,M}$} \ ,\\
 \lambda \frac{N^2(N-1)}{f}
  \big( \lambda e^{(N-1) X} + b \big)^{\frac{1}{f} - 1}
  e^{(N-1) \Psi} |\sigma|^{2N(N-1) - 2} \; ,
   \ \ \mbox{ for $SO(2N)/U(N)$} \ , \\
 \lambda \frac{N^2(N+1)}{h}
  \big( \lambda e^{(N+1) X} + b \big)^{\frac{1}{h} - 1}
  e^{(N+1) \Psi} |\sigma|^{2N(N+1) - 2} \; ,
   \ \  \mbox{ for $Sp(N)/U(N)$} \ ,
 \end{array}
 \right.
\end{align}
where $\Psi$ was defined in (\ref{Psi}).
These are singular at the surface $\sig=0$:
$g_{\sig\sig^*}|_{\sig=0}=0$.
This singularity is just a coordinate singularity of
$z^{\mu} =(\sig,\ph_{Aa})$.
To find a regular coordinate,
we perform coordinate transformations,
\begin{align}\label{rho}
 \rho \ &\equiv
\left\{
 \begin{array}{l}
  \sigma^{MN}/MN \; , \ \ \hs{13} \mbox{ for $G_{N,M}$} \ ,\\
  \sigma^{N(N-1)}/N(N-1) \; \ \ \mbox{ for $SO(2N)/U(N)$} \ , \\
  \sigma^{N(N+1)}/N(N+1) \; \ \ \mbox{ for $Sp(N)/U(N)$} \ ,
 \end{array}
 \right.
\end{align}
with $\ph_{Aa}$ (or $\ph_{ab}$) being unchanged.
The metrics in the regular coordinates
${z^{\mu}}' = (\rho,\ph_{Aa})$
can be calculated, to give
\bsubeq
\begin{align}
%%%%%%%% for G_{N,M} %%%%%%%%%%%%
g_{\rho \rho^*} \ &= \ \lambda \frac{M^2 N}{g}
 \big( \lambda e^{N X} + b \big)^{\frac{1}{g}-1} e^{N \Psi} \; , \\
g_{\rho (B b)^*} \ &= \ \lambda \frac{M^2 N^2}{g}
 \big( \lambda e^{N X} + b \big)^{\frac{1}{g} -1} e^{N \Psi}
 \rho^* \cdot \del_{(B b)^*} \Psi \; , \\
g_{(A a) (B b)^*} \ &= \ \lambda \frac{M^2 N^3}{g}
 \big( \lambda e^{N X} + b \big)^{\frac{1}{g} -1} e^{N \Psi}
 |\rho|^2 \cdot \del_{(A a)} \Psi \del_{(B b)^*} \Psi \nonumber \\
\ & \ \ \ \
 + \big( \lambda e^{N X} + b \big)^{\frac{1}{g}}
 \cdot \del_{(Aa)} \del_{(B b)^*} \Psi \; ,
\end{align}
\esubeq
for $G_{N,M}$,
%%%%%%%% for SO(2N)/U(N) %%%%%%%%%%%%
\bsubeq
\begin{align}
g_{\rho \rho^*} \ &= \ \lambda \frac{N^2(N-1)}{f}
 \big( \lambda e^{(N-1) X} + b \big)^{\frac{1}{f}-1} e^{(N-1) \Psi} \;,\\
g_{\rho (c d)^*} \ &= \ \lambda \frac{N^2 (N-1)^2}{f}
 \big( \lambda e^{(N -1) X} + b \big)^{\frac{1}{f} -1} e^{(N-1) \Psi}
 \rho^* \cdot \del_{(c d)^*} \Psi \; , \\
g_{(a b) (c d)^*} \ &= \ \lambda \frac{N^2 (N-1)^3}{f}
 \big( \lambda e^{(N-1) X} + b \big)^{\frac{1}{f} -1} e^{(N-1) \Psi}
 |\rho|^2 \cdot \del_{(a b)} \Psi \del_{(c d)^*} \Psi \nonumber \\
\ & \ \ \ \
 + \big( \lambda e^{(N-1) X} + b \big)^{\frac{1}{f}}
  \cdot \del_{(a b)} \del_{(c d)^*} \Psi \; ,
\end{align}
\esubeq
for $SO(2N)/U(N)$, and
%%%%%%%% for Sp(N)/U(N) %%%%%%%%%%%%
\bsubeq
\begin{align}
g_{\rho \rho^*} \ &= \ \lambda \frac{N^2(N+1)}{h}
 \big( \lambda e^{(N+1) X} + b \big)^{\frac{1}{h}-1}
 e^{(N+1) \Psi} \; , \\
g_{\rho (c d)^*} \ &= \ \lambda \frac{N^2 (N+1)^2}{h}
 \big( \lambda e^{(N +1) X} + b \big)^{\frac{1}{h} -1} e^{(N+1) \Psi}
 \rho^* \cdot \del_{(c d)^*} \Psi \; , \\
g_{(a b) (c d)^*} \ &= \ \lambda \frac{N^2 (N+1)^3}{h}
 \big( \lambda e^{(N+1) X} + b \big)^{\frac{1}{h} -1} e^{(N+1) \Psi}
 |\rho|^2 \cdot \del_{(a b)} \Psi \del_{(c d)^*} \Psi \nonumber \\
 \ & \ \ \ \ + \big( \lambda e^{(N+1) X} + b \big)^{\frac{1}{h}}
 \cdot \del_{(a b)} \del_{(c d)^*} \Psi \; ,
\end{align}
\esubeq
for $Sp(N)/U(N)$, where $\Psi$ differentiated by
the base coordinates $\ph_{Aa}$ or $\ph_{ab}$ are given in
(\ref{deriv-G}) and (\ref{deriv-SO-Sp}).

%%%
The metrics of the submanifold defined by
$\rho = 0$ ($d \rho = 0$) are
\bsubeq
\begin{align}
g_{(A a) (B b)^*}|_{\rho=0}\,(\ph,\ph^*)
 \ &= \ b^{\frac{1}{g}} \del_{(Aa)} \del_{(B b)^*} \Psi \;
   \ \ \mbox{ for $G_{N,M}$} \ ,\\
g_{(a b) (c d)^*}|_{\rho=0}\,(\ph,\ph^*)
 \ &= \ b^{\frac{1}{f}} \del_{(a b)} \del_{(c d)^*} \Psi \;
  \ \ \ \mbox{ for $SO(2N)/U(N)$} \ , \\
g_{(a b) (c d)^*}|_{\rho=0}\,(\ph,\ph^*)
 \ &= \ b^{\frac{1}{h}} \del_{(a b)} \del_{(c d)^*}\Psi \;
  \ \ \  \mbox{ for $Sp(N)/U(N)$} \ ,
\end{align}
\esubeq
where $\Psi$ is the \kahler potential of these manifolds
found in (\ref{Gra-Kahler}),
and $\Psi$ differentiated by two fields is given in
(\ref{2-deriv}) or (\ref{2-deriv-so-sp}).
Therefore we find that
the total spaces are the complex line bundles
over these Hermitian symmetric spaces as base manifolds,
with $\rho$ (or $\sig$) being a fiber.
Actually, it was shown in \cite{PP} that
there exists a Ricci-flat metric on the complex line bundle over
any K\"{a}hler-Einstein manifolds.
In the limit of $b\to 0$, these base manifolds shrink
and the manifolds become conifolds.
The conical singularities are resolved by
$G_{N,M}$, $SO(2N)/U(N)$ and $Sp(N)/U(N)$ of the radii $b^{1/2g}$,
$b^{1/2f}$ and $b^{1/2h}$, respectively.

%%%%%
~From the relation of $G_{N,1} = {\bf C}P^{N-1}$,
we also have the complex line bundle over ${\bf C}P^{N-1}$.
The \kahler potential (\ref{LG-kahler}) in the case of
$M=1$ ($g=N$) coincides with the flat one in the limit of $b\to 0$,
but with a coordinate identification $\rho = \sigma^N /N$.
The singular limit is ${\bf C}^N / {\bf Z}_N$,
and this orbifold singularity is resolved by ${\bf C}P^{N-1}$.
This coincides with the Calabi metric~\cite{Ca},
so our manifolds can be considered as natural generalizations
of the Calabi metric.

%%%%%%%%%%%%%%%%%%%%%%%%%%%%%%%%%%%%%%%%%%%%%%%
\section{Conclusion and Discussions}

We have constructed
non-compact \kahler manifolds,
modifying the \kahler quotient construction of
the Hermitian symmetric spaces of the classical groups
by restricting the gauge group $U(M)$ to $SU(M)$.
We have presented the Ricci-flat metrics and
their \kahler potentials on these manifolds.
The essential point was
that the partial differential equation (\ref{ricci-flat})
was reduced to the ordinary differential equations (\ref{ODE-G}),
(\ref{ODE-SO}) and (\ref{ODE-Sp}),
using the isotropy transformation.
These metrics contain the resolution parameter
as an integration constant,
and the conical singularities
are resolved by the Hermitian symmetric spaces
with the radii of the resolution parameter.
They have been recognized as
the line bundle over the Hermitian symmetric spaces,
and contain the Calabi metrics on
the line bundle over ${\bf C}P^{N-1}$.
Our manifolds in lower dimensions are discussed in Appendix,
which are not included in the list of \cite{CGLP}.

Our method can be applied to the cases
in which all non-compact directions
can be transformed to each other by the isotropy.
Such a view point was discussed in \cite{Ni1}
in terms of the supersymmetric nonlinear realization.
We can also construct other conifolds with the isometry of
the exceptional groups from
the Hermitian symmetric spaces of
the exceptional groups~\cite{HKN3}.
We would like to discuss whether the
deformation parameter exists, as in the case of
the conifold~\cite{HKN}.
We also would like to clarify the relation between
our manifolds of the line bundle over
$Sp(2)/U(2) \simeq Q^3$
and the Spin($7$) manifold in \cite{Spin7},
since both manifolds can be written in the form of
${\bf R}\times Sp(2)/SU(2)$.
The investigation of superconformal field theories
corresponding to our manifolds is also an interesting task.

%%%%%%%%%%%%%%%%%%%%%%%%%%%%%%%%%%%%%%%%%%%%%%%%%%%%%%%%%%%%%%%
\section*{Acknowledgements}
%%%%%%%%%%%%%%%%%%%%%%%%%%%%%%%%%%%%%%%%%%%%%%%%%%%%%%%%%%%%%%%
We would like to thank Michihiro Naka,
Kazutoshi Ohta and Takashi Yokono
for valuable comments and discussions.
This work was supported in part by the Grant-in-Aid for Scientific
Research (\#13640283).

%%%%%%%%%%%%%%%%%%%%%%%%
\begin{appendix}
\section{Isomorphism}
We have sets of the isomorphism between
the lower dimensional base manifolds,
\bsubeq
\begin{align}
{\rm i)}\hs{5}  &
  {\bf C}P^1 \ \simeq \ SO(4)/U(2)
  \ \simeq \ Sp(1)/U(1) \ \simeq \ Q^1 \; , \label{EHclass} \\
{\rm ii)}\hs{5}&
  {\bf C}P^3 \ \simeq \ SO(6)/U(3)\; , \label{Calabi4} \\
{\rm iii)}\hs{5} &
  Sp(2)/U(2) \ \simeq \ Q^3 \;, \label{CY4} \\
{\rm iv)}\hs{5} &
  G_{4,2} \ \simeq \ Q^4 \; , \label{CY5}
\end{align}
in addition to the novel duality relation
\begin{align}
{\rm v)}\hs{5} &
  G_{N,M} \ \simeq \ G_{N,N-M} \; . \label{dual-G}
\end{align}
\esubeq
In this appendix we show that total spaces
on these base manifolds coincide.
It gives a nontrivial check for our results.

Before doing that we quote the results of
the line bundle over $Q^{N-2}=SO(N)/[SO(N-2)\times U(1)]$,
as a conifold~\cite{HKN2}.
The superfields $\sig$ and $w_i$ ($i=1,\cdots N-2$)
constitute an $N$-vector as
$\Phi^T = \sig (1,w_i, -\1{2} \sum_{i=1}^{N-2}(w_i)^2)$.
The \kahler potential differentiated by
the invariant $X=\log \Phi\dagg\Phi$
and a coordinate transformation are
\begin{align}\label{QN}
 \Td{{\cal K}}{X}
   \ = \ (\lambda e^{(N-2)X} + b )^{\1{N-1}}  \; , \ \ \
  \sig \ = \ {\rho^{N-2} \over {N-2}} \; .
\end{align}
Note that the notation of $X$ is
different from that of \cite{HKN2}.

%%%%%%%%%%%%%%%%%%%%%%%%%%%%%%%%%%%%%%%%%%%%%%%%

%%%%%%%%%%%%%%%%%%%%%%%%%%%%%%%%%%%%%%%%%%%%%%%%%%%%
i) {\sl Eguchi-Hanson space.}
All of the lowest dimensional manifolds of (\ref{EHclass})
coincide with the Eguchi-Hanson gravitational instanton~\cite{EH}.
The matrix field $\Phi_I$ ($I=1,2,3,4$)
and invariants $X_I \equiv \log \det \Phi_I\dagg\Phi_I$ are
\bsubeq
\begin{align}
&\Phi_1 \ = \ \sigma_1
 \begin{pmatrix}
  1 \cr
  \varphi_1
\end{pmatrix} \; , \ \ \
X_1
 \ = \ \log |\sigma_1|^2 + \log ( 1 + |\varphi_1|^2 ) \; , \\
&\Phi_2 \
= \ \sigma_2
\begin{pmatrix}
      1 & 0 \cr
      0 & 1 \cr
      0 & \ph_2 \cr
 -\ph_2 & 0 \cr
\end{pmatrix}  \; , \ \ \
X_2
 \ = \ 2 \log |\sigma_2|^2 + 2 \log ( 1 + |\varphi_2|^2 ) \; , \\
&\Phi_3 \ = \ \sigma_3
\begin{pmatrix}
 1 \cr
 \varphi_3
\end{pmatrix} \; , \ \ \
X_3
 \ = \ \log |\sigma_3 |^2 + \log ( 1 + |\varphi_3|^2 ) \; , \\
&\Phi_4 \ = \ \sigma_4
\begin{pmatrix}
 1 \cr
 \ph_4 \cr
 - \half (\ph_4)^2
\end{pmatrix} \; , \ \ \
X_4
\ = \ \log |\sigma_4|^2 + 2 \log \Big( 1 + \half |\ph_4|^2 \Big) \; ,
\end{align}
\esubeq
for the cases of the base manifolds
${\bf C}P^1$, $SO(4)/U(2)$, $Sp(1)/U(1)$ and $Q^1$, respectively.
Relations of the base and the fiber coordinates of
these four manifolds are
$\varphi_1 = \varphi_2 = \varphi_3 = \ph_4/\sqrt{2}$
and $(\sigma_1)^2 \sim (\sigma_2)^2 \sim (\sigma_3)^2
\sim \sigma_4 \sim \rho$, respectively.
Note that each fiber $\sig_I$ $(I=1,2,3,4)$
consistently defines
the same $\rho$ as the regular coordinate from
(\ref{rho}) and (\ref{QN}).
The \kahler potentials coincide up to a overall constant:
\begin{align}
 {d {\cal K}\over d X_I}
 \ &\sim \ \sqrt{\lam e^{2X_1} + b}
 \ \sim \ \sqrt{\lam e^{X_2} + b}
 \ \sim \ \sqrt{\lam e^{2X_3} + b}
 \ \sim \ \sqrt{\lam e^{X_4} + b} \; .
\end{align}
The orbifold singularity in ${\bf C}^2/{\bf Z}_2$
is resolved by $S^2$.

%%%%%%%%%%%%%%%%%%%%%%%%%%%%%%%%%%%%%%%%%%%%%%%%%%%%
ii) {\sl Complex four-dimensional Calabi metric}. % (\ref{Calabi4}).
The matrix field $\Phi_I$ ($I=1,2$)
and invariants $X_I \equiv \log \det \Phi_I\dagg\Phi_I$ are
($i=1,2,3$)
\bsubeq
\begin{align}
&\Phi_1 \
= \ \sigma_1
\begin{pmatrix}
 1 \cr
 w_i
\end{pmatrix} \; , \ \ \
X_1 \ = \
\log |\sigma_1|^2
+ \log \Big( 1 + \sum_{i=1}^3 |w_i|^2 \Big) \; , \\
%%%
& \Phi_2 \ = \ \sigma_2
\left(
\begin{array}{ccc}
      1 & 0      & 0 \\
      0 & 1      & 0 \\
      0 & 0      & 1 \\
      0 & \ph_1  & \ph_2 \\
 -\ph_1 & 0      & \ph_3 \\
 -\ph_2 & -\ph_3 & 0
\end{array} \right) \; , \ \ \
X_2
\ = \ 3 \log |\sigma_2|^2
+ 2 \log \Big( 1 + \sum_{i=1}^3 |\ph_i|^2 \Big) \; ,
\end{align}
\esubeq
for the cases of the base manifolds ${\bf C}P^3$ and $SO(6)/U(3)$,
respectively.
Identifications of the base and the fiber coordinates are
$w_i = \varphi_i$ and
$(\sigma_1)^4 \sim (\sigma_2)^6 \sim \rho$,
respectively, where each fiber $\sig_I$ ($I=1,2$)
consistently defines
the same $\rho$ as the regular coordinate from (\ref{rho}).
The \kahler potentials coincide up to a overall constant:
\beq
 \Td{{\cal K}}{X_I}
 \ \sim \ (\lambda e^{4 X_1} + b)^{\1{4}}
 \ \sim \ (\lambda e^{2 X_2} + b)^{\1{4}} \;.
\eeq
The orbifold singularity in ${\bf C}^4/{\bf Z}_4$
is resolved by ${\bf C}P^3 \simeq SO(6)/U(3)$.

%%%%%%%%%%%%%%%%%%%%%%%%%%%%%%%%%%%%%%%%%%%%%%%%%%%%
iii) {\sl Another metric with complex four dimensions}.
The matrix field $\Phi_I$ ($I=1,2$)
and invariants $X_I \equiv \log \det \Phi_I\dagg\Phi_I$ are
($i=1,2,3$)
\bsubeq
\begin{align}
&\Phi_1 \ = \ \sigma_1
\begin{pmatrix}
                    1 & 0 \cr
                    0 & 1 \cr
                \ph_1 & {\ph_3 \over \sqrt 2} \cr
 {\ph_3\over \sqrt 2} & \ph_2
\end{pmatrix} \; , \ \ \
X_1
\ = \ 2 \log |\sigma_1|^2 +
\log \Big( 1 + \sum_{i=1}^3 |\varphi_i|^2
+ \Big|\1{2}\varphi_3^2 - \varphi_1 \varphi_2\Big|^2 \Big) \; , \\
%%%
& \Phi_2 \ = \ \sigma_2
\begin{pmatrix}
  1 \cr
  w_i \cr
 - \half \sum_{i=1}^3 w_i^2
\end{pmatrix} \; , \ \ \
X_2
\ = \ \log |\sigma_2|^2 + \log \Big( 1 + \sum_{i=1}^3 |w_i|^2 +
   \frac{1}{4} \Big| \sum_{i=1}^3 w_i^2 \Big|^2 \Big) \; ,
\end{align}
\esubeq
for the cases of the $Sp(2)/U(2)$ and $Q^3$ base manifolds.
Identifications of the base and the fiber coordinates are
\begin{align}
\left(
\begin{array}{c}
 \varphi_1 \\
 \varphi_2 \\
 \varphi_3 \\
\end{array} \right) \ &= \ \left(
\begin{array}{ccc}
 \frac{i}{\sqrt{2}} & - \frac{1}{\sqrt{2}} & 0\\
 \frac{i}{\sqrt{2}} & \frac{1}{\sqrt{2}}   & 0\\
                  0 & 0                    & 1
\end{array} \right) \left(
\begin{array}{c}
 w_1 \\
 w_2 \\
 w_3
\end{array} \right) \; ,
\end{align}
and $(\sigma_1)^6 \sim (\sigma_2)^3 \sim \rho$, respectively.
Again, each fiber $\sig_I$ $(I=1,2)$ consistently defines
the same $\rho$ as the regular coordinate from
(\ref{rho}) and (\ref{QN}).
The \kahler potentials coincide up to a overall constant:
\beq
 \Td{{\cal K}}{X_I}
 \ \sim \ (\lambda e^{3 X_1} + b)^{\1{4}}
 \ \sim \ (\lambda e^{3 X_2} + b)^{\1{4}} \; .
\eeq

%%%%%%%%%%%%%%%%%%%%%%%%%%%%%%%%%%%%%%%%%%%%%%%%%%%%
iv) {\sl The line bundle over the Klein quadric
(with complex five dimensions)}.
The embedding of $G_{4,2}$ into ${\bf C}P^5$ is known as
the Pl\"ucker embedding.
The matrix field $\Phi_I$ ($I=1,2$)
and invariants $X_I \equiv \log \det \Phi_I\dagg\Phi_I$ are
($i=1,2,3,4$)
\bsubeq
\begin{align}
&\Phi_1 \ = \ \sigma_1
\left(
\begin{array}{cc}
      1 & 0 \\
      0 & 1 \\
  \ph_1 & \ph_3 \\
  \ph_4 & \ph_2
\end{array} \right) \; , \ \ \
X_1 \ = 2 \log |\sigma_1|^2
 + \log \Big( 1 + \sum_{i=1}^4 |\ph_i|^2 +
 |\ph_1 \ph_2 - \ph_3 \ph_4|^2 \Big) \; , \\
%%%
& \Phi_2 \ = \ \sigma_2
\begin{pmatrix}
 1 \cr
 w_i \cr
 - \half \sum_{i=1}^4 w_i^2
\end{pmatrix} \; , \ \ \
 X_2 \ = \
 \log |\sigma_2|^2 + \log \Big( 1 + \sum_{i=1}^4 |w_i|^2
  + \frac{1}{4} \Big|\sum_{i=1}^4 w_i^2 \Big|^2 \Big) \; ,
\end{align}
\esubeq
for the cases of the base manifolds $G_{4,2}$ and $Q^4$.
Identifications of the base and the fiber coordinates are
\begin{align}
\left(
\begin{array}{c}
 \ph_1 \\
 \ph_2 \\
 \ph_3 \\
 \ph_4
\end{array} \right) \ &= \ \left(
\begin{array}{cccc}
\frac{1}{\sqrt{2}} & \frac{i}{\sqrt{2}} & 0 & 0 \\
\frac{1}{\sqrt{2}} & - \frac{i}{\sqrt{2}} & 0 & 0 \\
 0 & 0 & \frac{i}{\sqrt{2}} & - \frac{1}{\sqrt{2}} \\
 0 & 0 & \frac{i}{\sqrt{2}} & \frac{1}{\sqrt{2}}
\end{array} \right) \left(
\begin{array}{c}
 w_1 \\
 w_2 \\
 w_3 \\
 w_4
\end{array} \right) \; ,
\end{align}
and $(\sigma_1)^8 \sim (\sigma_2)^4 \sim \rho$,
respectively, from (\ref{rho}) and (\ref{QN}).
Each fiber defines the same the regular
fiber coordinate $\rho$.
The \kahler potentials coincide up to a overall constant:
\beq
 \Td{{\cal K}}{X_I}
\ \sim \ (\lambda e^{4 X_1} + b)^{\1{5}}
\ \sim \ (\lambda e^{4 X_2} + b)^{\1{5}} \, .
\eeq

%%%%%%%%%%%%%%%%%%%%%%%%%%%%%%%%%%%%%%%%%%%%%%%%%%%%
v) {\sl Duality between the Grassmann manifolds.}
The matrix field $\Phi_I$ ($I=1,2$)
and invariants $X_I \equiv \log \det \Phi_I\dagg\Phi_I$ are
\bsubeq
\begin{align}
&\Phi_1 \ = \ \sigma_1
\begin{pmatrix}
  {\bf 1}_{M} \cr
  \ph_1
\end{pmatrix} \; , \ \ \
X_1 \ = \ M \log |\sig_1|^2
 + \log \det ({\bf 1}_{M} + {\ph_1}^{\dagger} \ph_1) \; , \\
%%%
& \Phi_2 \ = \
 \sigma_2
\begin{pmatrix}
 {\bf 1}_{N-M} \cr
  \ph_2
\end{pmatrix} \; , \ \ \
X_2 \ = \  (N-M) \log |\sig_2|^2
  + \log \det ({\bf 1}_{N-M} + {\ph_2}^{\dagger} \ph_2) \; ,
\end{align}
\esubeq
for the cases of the base manifolds
$G_{N,M}$ and $G_{N,N-M}$, respectively.
Here $\ph_1$ and $\ph_2$ are [$(N-M) \times M$]-
and [$M \times (N-M)$]-matrices, respectively.
Identifications of the base and
the fiber coordinates are $\ph_1 = {\ph_2}^T$, and
$(\sig_1)^{NM} \sim (\sig_2)^{N(N-M)} \sim \rho$, respectively,
due to (\ref{rho}), in which each fiber defines
the same regular coordinate $\rho$.
The \kahler potentials coincide up to a overall constant:
\beq
 \Td{{\cal K}}{X_I}
\ \sim \ (\lambda e^{N X_1} + b)^{\1{g}}
\ \sim \ (\lambda e^{N X_2} + b)^{\1{g}} \, .
\eeq

\end{appendix}

%%%%%%%%%%%%%%%%%%%%%%%%%%%%%%%%%%%%%%%%%%%%%%%%%%%%%%%%%%%%%%%%%%%%
%%%%%%%%%%%%%%%%%%%% reference %%%%%%%%%%%%%%%%%%%%%%%%%%%%%%%%%%%%%
%%%%%%%%%%%%%%%%%%%%%%%%%%%%%%%%%%%%%%%%%%%%%%%%%%%%%%%%%%%%%%%%%%%%

\end{document}